\documentclass[12pt]{article}
\usepackage{cite, multicol, curves} 
\topmargin = - 8 mm
\newcommand{\eff}{{\overline{\alpha}}}
\newcommand{\Frac}[2]%
{{\textstyle \frac{\mbox{\footnotesize $#1$}\rule[-0.9mm]{0mm}{1mm}}%
{\mbox{\footnotesize $#2$}\rule{0mm}{3.1mm}}}}
\begin{document}
\vspace*{-12mm}
\begin{flushright}
hep-ph/9701321
\end{flushright}
\vskip 24 mm
\begin{center}
{\Large Relating inclusive $e^+e^-$ annihilation to 
electroproduction\\[2mm]
sum rules in Quantum Chromodynamics}\\[12mm]
{\large R. J. Crewther}\footnote{E-mail: rcrewthe@physics.adelaide.edu.au,
crewther@butp.unibe.ch}\\[5mm]
{\small\em Institute for Theoretical Physics, University of Berne,}\\
{\small\em Sidlerstrasse 5, CH-3012 Berne, Switzerland}\\
{\small\em and}\\
{\small\em Department of Physics and Mathematical Physics and}\\
{\small\em Centre for the Subatomic Structure of Matter,}\\
{\small\em University of Adelaide, Adelaide SA 5005, Australia}%
\footnote{Permanent address. On leave of absence until February, 1997.}
\end{center}
\vskip 2mm
\begin{abstract}\noindent
The Broadhurst-Kataev conjecture, that the ``discrepancy'' 
in the connection with the $\pi^0 \rightarrow \gamma\gamma$ anomaly
equals the beta function $\beta({\eff})$ times a power 
series in the effective coupling $\eff $, is proven to all orders of 
perturbative quantum chromodynamics. The use of nested short-distance 
expansions is justified via Weinberg's power-counting theorem.
\end{abstract}
\vspace{14mm}

There has been a revival of interest \cite{bk,brod1,brod2,kat}
in the relation \cite{rjc}
\begin{equation}  
3S = KR'
\label{a1}\end{equation}
between the anomalous constant $S$ \cite{BJ,adler1} governing 
$\pi^0 \rightarrow \gamma\gamma$ decay, the isovector part $R'$ 
of the cross-section 
ratio
\begin{equation}
R = \{e^+e^- \rightarrow \mbox{hadrons}\}/
     \{e^+e^- \rightarrow \mu^+\mu^-\}
\label{a2}\end{equation}
at large centre-of-mass energy $\sqrt{s}/2$, and the lowest 
isovector moment \cite{bj} of the first spin-dependent 
structure function $g_1(x,Q^2)$ for inclusive electroproduction 
at large momentum transfer $Q$: 
\begin{equation}
\int^1_0\! dx\, g^{ep-en}_1(x,Q^2) = 
\frac{1}{6}\left|\frac{g_A}{g_V}\right|K(Q^2) + O(Q^{-2})
\label{a3}\end{equation}
\newpage\noindent
The result (\ref{a1}) was derived
in a non-perturbative fashion before the advent of quantum 
chromodynamics (QCD), with the hadronic energy-momentum tensor 
$\theta_{\mu\nu}$ assumed to have a soft trace:%
\footnote{In effect, conformal invariance was supposed to be softly
broken and so valid for leading powers at short distances. The 
immediate precursors of \cite{rjc} were papers by Schreier 
\cite{schreier} and Migdal \cite{migdal}. Some conformal aspects 
of \cite{rjc} were found independently by Ferrara et al.\ 
\cite{ferrara}.}

\begin{equation}
\mbox{dim}\,\theta^\mu_\mu < 4\ ,\ \ \mbox{pre-QCD}  
\label{a4}\end{equation}
Of course, there is a trace anomaly \cite{rjc,ce} in QCD \cite{trace} 
which violates (\ref{a4}). Nevertheless, the relation (\ref{a1}) 
remains valid for QCD in leading logarithm approximation:
$K \rightarrow 1$ and $R' \rightarrow N_c/2$, where $N_c$ is the
number of colours. The choice $N_c=3$ fits the observed value 
$S_{\rm expt} \simeq 0.5$.

Broadhurst and Kataev \cite{bk} tried extending (\ref{a1}) to the
leading QCD \emph{power}, i.e.\ to corrections like
$K(Q^2)$ in (\ref{a3}) and $D(Q^2)$ in the Adler function 
\cite{adler2} 
\begin{equation}
Q^2\int^\infty_{4m_\pi^2}\! ds\, R(s)/(s+Q^2)^2
= D(Q^2)R(\infty) + O(Q^{-2})
\label{b1}\end{equation}
The axial-vector current was required to be flavour 
\emph{non-singlet\/} as in \cite{rjc}, but the vector currents 
could be either flavour singlet ($\mathcal{F} = S$) or non-singlet 
($\mathcal{F} = NS$). Products $K_{\!\mathcal{F}}D_{\!\mathcal{F}}$
were tested using existing multi-loop results \cite{K,D} for
the $\mathcal{F} = S,NS$ versions of $K$ and $D$. Here $D_{NS}$ and 
$D_{S}$ are the factors in (\ref{b1}) produced by the isovector and
baryon currents, $K_{NS}$ is the factor $K(Q^2)$ in the Bjorken sum 
rule (\ref{a3}), and $K_{S}$ gives the leading power in the sum 
rule of Gross and Llewellyn Smith \cite{gls} (but \emph{not\/} 
that of Ellis and Jaffe \cite{ej}).

Also considered \cite{bk} were the $Q$-independent corrections 
$K^*$ and $D^*$ due to subsets of Feynman diagrams in which the 
gauge coupling $g$ is not renormalized. Actually, the comparison was 
with an Abelian calculation \cite{gkls}, but one can imagine a 
conformal non-Abelian extension obtained by neglecting self energies 
in an axial gauge.

Broadhurst and Kataev \cite{bk} found, in agreement with early work 
of Adler, Callan, Gross and Jackiw \cite{acgj}, that $K^*$ and $D^*$ 
satisfy (\ref{a1}) to the highest order of calculation available, i.e.
\begin{equation}
K^*_{\!\mathcal{F}}(\alpha_s)
D^*_{\!\mathcal{F}}(\alpha_s)
= 1\ , \ \ \mathcal{F} = S,NS 
\label{b2}\end{equation}
up to terms  $O(\alpha_s^4)$ in $\alpha_s = g^2/4\pi$. With all 
diagrams included, $K$ and $D$ become power series in the effective 
coupling constant $\eff = \eff(Q^2/\mu^2,\alpha_s)$ defined by
\begin{equation}
\log(Q^2/\mu^2) = \int^\eff_{\alpha_s}\! d\alpha'/\beta(\alpha')
\label{b3}\end{equation}
where $\mu$ denotes a suitable renormalization scale. Then Broadhurst
and Kataev found that the relation (\ref{a1}) is violated by a term
proportional to $\beta(\eff)$,
\begin{equation}
K_{\!\mathcal{F}}(\eff)D_{\!\mathcal{F}}(\eff) 
= 1 + \beta(\eff)\times\{\mbox{power series in $\eff$}\},\
\ \ \mathcal{F} = S,NS
\label{b4}\end{equation}
again to the $O(\eff^4)$ accuracy of current calculations. They
christened the extra term ``the Crewther discrepancy''. 

In this letter, equations (\ref{b2}) and (\ref{b4}) will be shown 
to hold to all orders of perturbation theory. (An independent derivation
of (\ref{b4}) by D. M\"{u}ller, which I have not seen, is expected in
the near future \cite{dm}.) The derivation runs along the lines of 
\cite{rjc}, except that the pre-QCD assumption (\ref{a4}) is replaced 
by an analysis based on renormalized conformal Ward identities 
\cite{conf} and the QCD trace anomaly 
\cite{trace}
\begin{equation}
\theta^\mu_\mu = \frac{\beta(\alpha_s)}{\alpha_s}F^2 
+ \mbox{quark-mass\ terms}
\label{c1}\end{equation}
Here $F^2$ denotes the renormalized square of the gluonic field strength 
tensor $F^a_{\mu\nu}$.

Let $J_\mu(x)$ and $J_{\mu5}(x)$ be the electromagnetic current and the
isovector axial-vector current in QCD. As an operator on hadron states, 
$J_{\mu5}$ is almost conserved: its divergence 
$\partial^\mu J_{\mu5} = \Delta(x)$ is proportional to the light-quark 
masses $m_u,m_d$. In any order of perturbation theory, $\Delta$ carries
a  dynamical dimension of 3 modified by QCD logarithms. 
The condition $\mbox{dim}\,\Delta < 4$ is satisfied, so Wilson's 
prescription \cite{kgw} for the anomalous constant $S$ can be adopted 
without change:%
\footnote{Perturbatively, photon channels have no mass gap because 
of the three-gluon threshold (for example), but the effect is infra-red 
finite and hence not a problem. In the pseudoscalar channel, the light 
$q\bar{q}$ pair provides a factor $\sim (m_u + m_d)^{-1}$ \cite{gth}
which cancels the explicit mass dependence of
$\Delta$ and so produces a result for $S$ independent of $m_u$ and
$m_d$.}
\begin{equation}
S = \frac{\pi^2}{12}\,\varepsilon^{\mu\nu\alpha\beta}\!
\int\hspace{-2mm}\int_{{\cal R}_\epsilon}\! d^4x\, d^4y\, x_\mu y_\nu
T\langle\mbox{vac}|J_\alpha(x)J_\beta(0)\Delta(y)|\mbox{vac}\rangle
+ O(\epsilon)
\label{c2}\end{equation}
Here ${\cal R}_\epsilon$ is the entire eight-dimensional volume, except
that narrow regions of width $O(\epsilon)$ containing coincident points
$x_\mu=0$, $y_\mu=0$, or $x_\mu=y_\mu$ are excluded (Fig.~\ref{one}). By
definition, S remains independent of $\epsilon$.

\begin{figure*}
\setlength{\columnsep}{9mm}
\begin{multicols}{2}
\setlength{\unitlength}{1mm}
\begin{picture}(68,49)
\put(9.1,27){\vector(0,1){2}}
\put(9.1,27){\vector(0,-1){2}}
\put(14,19){${\cal R}_\epsilon$}
\put(47.4,37.4){${\cal S}_\epsilon$}
\put(1,26.3){\footnotesize $O(\epsilon)$}
\put(57,26.3){\footnotesize $y_0 = 0$}
\put(29,1){\footnotesize $x_0=0$}
\put(5,3){\footnotesize $x_0=y_0$}
\put(34,3.5){\vector(0,1){5}}
\put(13,6){\vector(1,1){3.6}}
\put(41,16){\sffamily\footnotesize ANOM}
\put(15,41.5){\sffamily\footnotesize ETC}
\put(23.5,43){\vector(1,0){7}}
\put(17.5,39.5){\vector(0,-1){8.5}}
\put(41,20){\vector(-1,1){4}}
\linethickness{0.28mm}
\put(11,27){\curve(0,0, 44.5,0)}
\put(34,10){\curve(0,0, 0,38.5)}
\put(18,11){\curve(0,0, 35,35)}
\curvedashes[1mm]{0,1.2,1.8}
\put(36,32){\curve(0,0, 0,17)}
\put(36,31.8284){\curve(0,0, 15,15)}
\put(39,29){\curve(0,0, 20,0)}
\put(38.8284,29){\curve(0,0, 8,8)}
\put(51.5562,41.7278){\curve(0,0, 4,4)}
\put(36,25){\curve(0,0, 0,-19)}
\put(36,25){\curve(0,0, 23,0)}
\put(32,22){\curve(0,0, 0,-16)}
\put(32,22.1716){\curve(0,0, -15.5,-15.5)}
\put(29,25){\curve(0,0, -18.6,0)}
\put(29.1716,25){\curve(0,0, -15.5,-15.5)}
\put(32,29){\curve(0,0, -21.6,0)}
\put(32,29){\curve(0,0, 0,20)}

\end{picture}

\caption{\label{one}
A volume of integration ${\cal R}_\epsilon$ in $(x,y)$ 
space \cite{kgw} bounded by a surface ${\cal S}_\epsilon$ (dashed 
lines); see (\ref{c2}) and (\ref{c4}). Regions 
{\sffamily\scriptsize ETC} where equal-time commutators may form do 
not contribute as $\epsilon \rightarrow 0$. The anomaly comes from 
the $x,y \rightarrow 0$ region {\sffamily\scriptsize ANOM}.}
\setlength{\unitlength}{1mm}
\begin{picture}(68,49)
\put(9,2){$J_\alpha$}
\put(55,2){$J_\beta$}
\put(32,46){$J_{\gamma 5}$}
\put(15,5){\circle*{1}}
\put(53,5){\circle*{1}}
\put(34,43){\circle*{1}}
\put(25,25){\circle*{1}}
\put(30.6568,19.3432){\circle*{1}}
\put(34,12.5){\circle*{1}}
\put(36.6926,5){\circle*{1}}
\linethickness{0.28mm}
\put(15,5){\curve(0,0, 38,0)}
\put(15,5){\curve(0,0, 19,38)}
\put(53,5){\curve(0,0, -19,38)}
\newcommand{\head}{\curve(0,0, 0.8,0.4, 1.5,1)%
\curve(0,0, 0.8,-0.4, 1.5,-1)}
\newcommand{\rhead}{\curve(0,0, -0.8,-0.4, -1.5,-1)%
\curve(0,0, -0.8,0.4, -1.5,1)}
\put(25,5){\head}
\put(44.5,5){\head}
\renewcommand{\xscale}{0.43896}
\renewcommand{\xscaley}{0.8985}
\renewcommand{\yscalex}{-0.8985}
\renewcommand{\yscale}{0.43896}
\put(30.6568,19.3432){\closecurve(0,0, 3.8081,3, 7.6162,0, 3.8081,-3)}
\put(34.7,17.8){\head}
\put(30,14){\rhead}
\newcommand{\glue}%
{\curve(0,0, 0.5,0.6, 1,0)\curve(1,0, 1.5,-0.6, 2,0)
\curve(2,0, 2.5,0.6, 3,0)\curve(3,0, 3.5,-0.6, 4,0)
\curve(4,0, 4.5,0.6, 5,0)\curve(5,0, 5.5,-0.6, 6,0)
\curve(6,0, 6.5,0.6, 7,0)\curve(7,0, 7.5,-0.6, 8,0)}
\renewcommand{\xscale}{0.7071}
\renewcommand{\xscaley}{0.7071}
\renewcommand{\yscalex}{-0.7071}
\renewcommand{\yscale}{0.7071}
\put(25,25){\glue}
\renewcommand{\xscale}{-0.3368}
\renewcommand{\xscaley}{-0.9375}
\renewcommand{\yscalex}{0.9375}
\renewcommand{\yscale}{-0.3368}
\put(36.6926,5){\glue}
\renewcommand{\xscale}{0.4472}
\renewcommand{\xscaley}{-0.8944}
\renewcommand{\yscalex}{0.8944}
\renewcommand{\yscale}{0.4472}
\put(20.5,16){\rhead}
\put(30,35){\rhead}
\renewcommand{\xscale}{0.4472}
\renewcommand{\xscaley}{0.8944}
\renewcommand{\yscalex}{-0.8944}
\renewcommand{\yscale}{0.4472}
\put(44,23){\rhead}
\renewcommand{\xscale}{1}
\renewcommand{\xscaley}{0}
\renewcommand{\yscalex}{0}
\renewcommand{\yscale}{1}
\end{picture}
\caption{\label{two}
Example of internal coupling constant renormalization for 
the VVA amplitude. Three-loop contributions to the leading power 
$C_{\alpha\beta\gamma}$ in (\ref{c3}) scale with degree $-9$ 
in $x$ and $y$ (like the bare triangle) but are not conformal 
invariant \cite{acgj}.}
\end{multicols}
\end{figure*}

Wilson noticed that (\ref{c2}) is the volume integral of an 
eight-divergence which can be converted into an integral of the 
VVA amplitude $T\langle J_\alpha J_\beta J_{\gamma 5}\rangle$
over the short-distance surface ${\cal S}_\epsilon$ bounding
${\cal R}_\epsilon$. Only the leading power 
$C_{\alpha\beta\gamma}$ multiplying the identity operator $I$ in 
the expansion
\begin{equation}
T\{J_\alpha(x)J_\beta(0)J_{\gamma 5}(y)\} 
\sim C_{\alpha\beta\gamma}(x,y)I + \ldots\;, \ \ x,y \rightarrow 0
\label{c3}\end{equation}
can contribute in the limit $\epsilon \rightarrow 0$:
\begin{eqnarray}
S &=& \frac{\pi^2}{12}\,\varepsilon^{\mu\nu\alpha\beta}\!
\int\hspace{-2mm}\int_{{\cal S}_\epsilon}\! \Big\{d^4x\, dS^\gamma_y\, 
x_\mu (y_\nu C_{\alpha\beta\gamma} - y_\beta C_{\alpha\gamma\nu})
\nonumber \\
& & \hphantom{\frac{\pi^2}{12}\,\varepsilon^{\mu\nu\alpha\beta}\!
\int\hspace{-2mm}\int_{{\cal S}_\epsilon}} + dS^\gamma_x\, d^4y\, 
y_\beta (x_\alpha C_{\gamma\mu\nu} - x_\mu C_{\alpha\gamma\nu})
\Big\} + O(\epsilon)
\label{c4}\end{eqnarray}
In QCD perturbation theory, $C_{\alpha\beta\gamma}$ consists of:
\begin{enumerate}
\item a lowest-order contribution $S\Delta_{\alpha\beta\gamma}(x,y)$
from the two bare triangle diagrams. Schreier \cite{schreier} 
showed that the $x,y$ dependence of $\Delta_{\alpha\beta\gamma}$ is 
allowed uniquely by conformal invariance. The resulting integral in 
(\ref{c4}) was performed as part of the non-perturbative analysis of 
\cite{rjc} and produced the expected answer.
\item amplitudes which break conformal invariance as a result of 
internal coupling constant renormalization. The simplest example is 
shown in Fig.~\ref{two}. The Adler-Bardeen theorem 
\cite{adler1,ab,bard,shei} requires that these amplitudes do not 
contribute to $S$, even though some contributions at four loops and 
beyond are logarithmically more singular at $x,y \sim 0$ than the 
bare triangle diagrams.
\end{enumerate}

In \cite{rjc}, the asymptotic three-point amplitude 
$C_{\alpha\beta\gamma}$ was analysed by substituting an expansion
of the form
\begin{equation}
T\{J_\alpha(x)J_\beta(0)\} 
= C^R_{\alpha\beta}(x)I + C_{\alpha\beta}^{K\:\mu}(x)J_{\mu 5}(0) 
 + \ldots\;, \ \ x_\mu \rightarrow 0
\label{d1}\end{equation}
in $T\{J_\alpha J_\beta J_{\gamma 5}\}$ and then expanding as follows:
\begin{equation}
T\{J_{\mu 5}(0)J_{\gamma 5}(y)\} 
= C^{R'}_{\mu\gamma}(y)I + \ldots\;, \ \ y_\mu \rightarrow 0
\label{d2}\end{equation}
Matching the expansions (\ref{c3}), (\ref{d1}) and (\ref{d2}) gave the
constraint
\begin{equation}
C_{\alpha\beta\gamma}(x,y) \rightarrow 
C^{K\:\mu}_{\alpha\beta}(x)C^{R'}_{\mu\gamma}(y)\ , \ \ x \ll y
\label{d3}\end{equation}
The superscripts $K$ and $R'$ label amplitudes related to the Bjorken
sum rule (\ref{a3}) and the isovector part of (\ref{a2}). (The limit\
$y \ll x$, where $T\{J_\beta(0)J_{\gamma 5}(y)\}$ is expanded
first, produces the connection with the Gross-Llewellyn Smith sum 
rule \cite{gls}.)

The technique leading to (\ref{d3}) depends on an interchange of 
{\em nested\/} short distance limits $\rho_1 \rightarrow 0$ and
$\rho_2 \rightarrow 0$ for operator products
\begin{equation}
T\Big\{\prod_i\{A_i(\rho_1\rho_2\hat{x}_i)\}B(0)
\prod_j\{C_j(\rho_2\hat{y}_j)\}\Big\}
\label{d4}\end{equation}
with $\hat{x}_i$ and $\hat{y}_j$ held fixed. Unlike most limit interchanges 
at short-distances, this {\em always\/} works, in any order of renormalized 
perturbation theory. The required uniformity property is contained in 
Weinberg's power-counting theorem \cite{wein}.

Weinberg introduced a $4N$-dimensional vector $\mathbf{P}$ to describe
the asymptotic behaviour of Euclidean amplitudes depending on $N$
four-dimensional momenta:
\begin{equation}
\mathbf{P} = \mathbf{L}_1\eta_1\eta_2\ldots\eta_m 
             +\mathbf{L}_2\eta_2\ldots\eta_m + \ldots 
             + \mathbf{L}_m\eta_m + \mathbf{C}
\label{d5}\end{equation}
Here $\eta_1 \ldots \eta_m$ are large positive parameters
corresponding to an arbitrary set of $m$ independent fixed 
vectors $\mathbf{L}_1,\ldots,\mathbf{L}_m\ (m \leq 4N)$, and 
$\mathbf{C}$ is a bounded vector. The theorem states that all 
amplitudes belong to asymptotic classes labelled by characteristic 
asymptotic powers (and powers of logarithms \cite{fink}) as 
$\eta_1\ldots\eta_m$ tend {\em independently\/} to infinity. In 
other words, the $\eta_i \rightarrow \infty$ limits are uniform with 
respect to each other: they can be carried out in any order.

Evidently
\begin{equation}
\eta_1 = \rho_1^{-1}\ ,\ \ \eta_2 = \rho_2^{-1}
\end{equation}
are special cases of Weinberg's asymptotic $\eta$ parameters. For any
Green's function containing (\ref{d4}) as a sub-product, we can 
subtract off terms in
\begin{equation}
T\Big\{\prod_i\{A_i(\rho_1\rho_2\hat{x}_i)\}B(0)\Big\}
\sim \sum_m {\cal C}_m(\rho_1\rho_2,\{\hat{x}_i\}){O}^\prime_m(0)\ ,
\ \ \rho_1 \rightarrow 0
\end{equation}
and
\begin{equation}
T\Big\{{O}^\prime_m(0)\prod_j\{C_j(\rho_2\hat{y}_j)\}\Big\}
\sim \sum_n {\cal C}_{mn}(\rho_2,\{\hat{y}_j\})O_m(0)\ ,
\ \ \rho_2 \rightarrow 0
\end{equation}
to remove as many asymptotic powers in $\rho_1$ and $\rho_2$ as we 
wish, and be sure that, relative to (\ref{d4}) in Euclidean space, the 
remainder 
\begin{equation}
\mathcal{R}_{MN} = T\{\prod_i A_i\, B \prod_j C_j\} 
    - \sum_{m=1}^M\sum_{n=1}^N {\cal C}_{mn}{\cal C}_nO_n
\end{equation}
will be $O(\rho_1^{-M}\rho_2^{-M-N}\times
\{\mbox{logs of $\rho_1$ and $\rho_2$\}})$ as $\rho_1$ and
$\rho_2$ tend independently to 0. Therefore the coefficient functions 
$f_n$ in the expansion $\sum_n f_nO_n$ of (\ref{d4}) for 
$\rho_2 \rightarrow 0$ must obey the rule \cite{rjc}
\begin{equation}
f_n \sim \sum_n {\cal C}_m{\cal C}_{mn}\ ,\ \ \rho_1 \rightarrow 0
\end{equation}
The interchange of the limits $\rho_1 \rightarrow 0$ and 
$\rho_2 \rightarrow 0$ to obtain conditions such as (\ref{d3}) is
thus justified. Note that conformal invariance is {\em not\/}
assumed --- the result is absolutely general.%
\footnote{This should have been discussed in 1972. At the time, I tried
to find a natural extension of axiomatic field theory to incorporate
Weinberg's asymptotic classes, but the project proved to be too 
ambitious --- hence the mysterious reference 10 in \cite{rjc} which
never appeared.}

For conformal subsets of graphs, the analysis of \cite{rjc} is fully
applicable, with $R'$ and $K$ replaced by 
$\frac{1}{2}N_c D^*_{\cal F}(\alpha_s)$ and $K^*_{\cal F}(\alpha_s)$.
Hence equation (\ref{b2}) is valid to all orders of perturbation
theory. 

In general, internal coupling constant renormalization breaks 
conformal invariance in leading powers and causes logarithms of 
$\rho_1$ and $\rho_2$ to appear. This occurs first for three-loop 
diagrams (Fig.~\ref{two}):
\begin{equation}
T\langle J_\alpha(\rho_1\rho_2\hat{x})J_\beta(0)
   J_{\gamma 5}(\rho_2\hat{y})\rangle_{\mbox{3-loop}}
= O(\rho_2^{-9}\rho_1^{-3}\log\rho_1)
\end{equation}
Effects of this type are controlled by conformal Ward identities
\cite{conf} in which the current
\begin{equation}
{\cal K}_{\mu\nu} 
= (2x^\lambda x_\nu - \delta^\lambda_\nu x^2)\theta_{\mu\lambda}(x)
\end{equation}
has a divergence containing the QCD trace anomaly \cite{trace}:
\begin{equation}
\partial^\mu{\cal K}_{\mu\nu} 
= 2x_\nu\frac{\beta(\alpha_s)}{\alpha_s}F^2 + \mbox{quark-mass\ terms}
\end{equation}

Let ${\cal D}_\nu$ denote the matrix differential operator which 
induces an infinitesimal conformal transformation on a current:
\begin{equation}
{\cal D}_\nu J_\alpha(x)
= 2x_\nu(3 + x\cdot\partial)J_\alpha - x^2\partial_\nu J_\alpha
  + 2g_{\nu\alpha}x\cdot J - 2x_\alpha J_\nu
\end{equation}
In order to write down the conformal Ward identity for the VVA
amplitude, it is necessary to have ${\cal D}_\nu$ act separately on
$J_\alpha(x)$, $J_\beta(x')$ (with $x'$ temporarily not set to zero) and
$J_{\gamma 5}(y)$, and then take the sum. The result is a first-order 
differential equation in $x,x',y$ space:
\begin{eqnarray}
\lefteqn{\Bigl(\sum_{\rm currents} {\cal D}_\nu\Bigr) 
T\langle J_\alpha(x)J_\beta(x')J_{\gamma 5}(y)\rangle} \nonumber \\
&=& \frac{\beta(\alpha_s)}{\alpha_s}
  T\langle J_\alpha J_\beta J_{\gamma 5}
  \int\!\! d^4z\, 2z_\nu F^2(z)\rangle
  + \mbox{non-leading\ powers}
\label{f1}\end{eqnarray}
For the moment, regions in which $x,x',y$ coincide are excluded.

The solution of (\ref{f1}) consists of a particular integral and a
homogeneous part. Schreier's work \cite{schreier} requires the
homogeneous part to be proportional to the bare triangle amplitude
$\Delta_{\alpha\beta\gamma}$. So, setting $x'=0$ once more, we find
that the leading power $C_{\alpha\beta\gamma}$ in (\ref{c3}) is given
by
\begin{equation}
C_{\alpha\beta\gamma} 
= c\,\Delta_{\alpha\beta\gamma}(x,y)
 + \beta(\alpha_s)\overline{\Delta}_{\alpha\beta\gamma}(x,y;\alpha_s)
\label{f2}\end{equation}
where $\overline{\Delta}_{\alpha\beta\gamma}$ is a power series in
$\alpha_s$ and $c$ is a constant of integration.

Equation (\ref{f2}) can be substituted into the surface integral
(\ref{c4}), with the result
\begin{equation}
S = c + \beta(\alpha_s)\overline{S}(\alpha_s)
\end{equation}
where $\overline{S}$ is the contribution due to 
$\overline{\Delta}_{\alpha\beta\gamma}$. This allows us to eliminate
$c$ from (\ref{f2}):
\begin{equation}
C_{\alpha\beta\gamma} 
= S\Delta_{\alpha\beta\gamma}(x,y)
+ \beta(\alpha_s)\{\overline{\Delta}_{\alpha\beta\gamma}(x,y;\alpha_s)
  - \overline{S}(\alpha_s)\Delta_{\alpha\beta\gamma}(x,y)\}
\end{equation}

The next step is to substitute the $x \ll y$ constraint (\ref{d3}).
This is less straightforward than in \cite{rjc}: unlike (\ref{a1}),
the desired result (\ref{b4}) contains momentum dependent amplitudes
which include contributions from coincident points $x=0=y$. At such
points, operator product expansions need not be generally valid
because of renormalization ambiguities proportional to $\delta^4$
functions and their derivatives \cite{cargese}. In this case, the
problem is solved by making a second use of electromagnetic gauge
invariance (the first being Wilson's prescription (\ref{c2})), and
projecting out the Adler function.

Since $C_{\alpha\beta\gamma}$, $\Delta_{\alpha\beta\gamma}$
and $\overline{\Delta}_{\alpha\beta\gamma}$ are superficially
linearly divergent, they may have ambiguities linear in momenta, 
i.e.\ proportional to $(\partial_x\ \mbox{or}\
\partial_y)\delta^4(x)\delta^4(y)$ in coordinate space. As is well
known, these ambiguities can be removed by imposing electromagnetic
gauge invariance as renormalization conditions of the form
\begin{equation}
\partial_x^\alpha C_{\alpha\beta\gamma} \ =\ 0\ 
=\ (\partial_x + \partial_y)^\beta C_{\alpha\beta\gamma} 
\end{equation}
The resulting amplitudes are then defined uniquely for all $x,y$,
with non-canonical (anomalous) results for
$\partial^\gamma_yC_{\alpha\beta\gamma}$ and 
$\partial^\gamma_y\Delta_{\alpha\beta\gamma}$. (Note that Lorentz 
covariance is also imposed as a {\em renormalization\/} condition; 
canonical constructions such as $T^*$-products should be avoided.) 

With $C_{\alpha\beta\gamma}$ thus well defined, its $x \ll y$ limit
will also respect electromagnetic gauge invariance. There is no problem
with $C_{\alpha\beta}^{K\:\mu}(x)$: it converges superficially, so it
can be extended to $x=0$ without ambiguity, with current conservation
maintained for the indices $\alpha$ and $\beta$. However the amplitude
$C^{R'}_{\mu\gamma}(y)$ has a superficial quadratic divergence.
Electromagnetic gauge invariance reduces this to a superficial
logarithmic divergence,
\begin{equation}
C^{R'}_{\mu\gamma}(y) 
= (g_{\mu\gamma}\partial^2 - \partial_\mu\partial_\gamma)\Pi^{R'}(y)
\end{equation}
but does not specify the subtraction procedure which fixes the 
$\delta^4(y)$ term in $\Pi^{R'}(y)$. The actual subtraction procedure
is determined implicitly via the $x \ll y$ limit of 
$C_{\alpha\beta\gamma}$; presumably it has a complicated $\alpha_s$
dependence, so this information is not very useful.

Fortunately, we do not need this information. One of the advantages of
the Adler function is that it does not depend on the subtraction
procedure used to renormalize the hadronic vacuum polarization
\cite{adler2}. In the leading power
\begin{equation}
D_{NS}(-q^2)R'(\infty) 
= \Frac{1}{2}q\cdot\frac{\partial}{\partial q} 
  \int\! d^4y\, e^{iq\cdot y}\Pi^{R'}(y)
\end{equation}
the ambiguity in $\Pi^{R'}$ is a constant in momentum space which is
removed by the operator $q\cdot\partial/\partial q$. In coordinate
space, this corresponds to
\begin{equation}
-{\textstyle \frac{1}{2}}(6 + y\cdot\partial)C^{R'}_{\mu\gamma}(y) 
= (g_{\mu\gamma}\partial^2 - \partial_\mu\partial_\gamma)\partial^\lambda
\{-{\textstyle \frac{1}{2}}y_\lambda \Pi^{R'}(y)\}
\end{equation}
where the factor $y_\lambda$ eliminates any term in $\Pi^{R'}(y)$
proportional to $\delta^4(y)$.

Therefore, all ambiguities at coinciding points can be eliminated by
imposing electromagnetic gauge invariance and considering
$(6 + y\cdot \partial/\partial y)C_{\alpha\beta\gamma}$ for 
$x \ll y$. A similar  analysis can be performed at $y \ll x$ to obtain 
the singlet case. Going to momentum space, we have
\begin{eqnarray}
\lefteqn{K_{\cal F}(\eff(P^2/\mu^2,\alpha_s))
D_{\cal F}(\eff(Q^2/\mu^2,\alpha_s))}   \nonumber \\
&=& 1 + \beta(\alpha_s)\overline{K}_{\cal F}(P^2,\alpha_s)
               \overline{D}_{\cal F}(Q^2,\alpha_s)\ ,
\ \ {\cal F} = S,NS
\end{eqnarray}
where $\overline{K}_{\cal F}$ and $\overline{D}_{\cal F}$ are power
series in $\alpha_s$.

The final step is to set $P=Q$ and substitute 
\begin{equation}
\beta(\alpha_s) = \beta(\eff)(\partial\eff/\partial\alpha_s)^{-1}
\end{equation}
to obtain
\begin{eqnarray}
K_{\cal F}(\eff)D_{\cal F}(\eff) 
&=& 1 + \beta(\eff){\cal P}_{\cal F}(Q^2,\alpha_s) \label{g1} \\
{\cal P}_{\cal F} &=& \overline{K}_{\cal F}(Q^2,\alpha_s)
\overline{D}_{\cal F}(Q^2,\alpha_s)(\partial\eff/\partial\alpha_s)^{-1}
\end{eqnarray}
for ${\cal F} = S,NS$. Since ${\cal P}_{\cal F}$ is a power series in
$\alpha_s$ and the rest of (\ref{g1}) depends only on $\eff$, we have
\begin{equation}
\mathcal{P}_\mathcal{F} = \mbox{power\ series\ in\ $\eff$} 
\end{equation}
Hence equation (\ref{b4}) is proven to all orders in perturbation theory.

It has been observed \cite{brod2} that the first two terms of 
$\mathcal{P}_\mathcal{F}$ can be removed by using different commensurate
scales $Q,Q^*$ for $K$ and $D$:
\begin{equation}
K_\mathcal{F}(Q)D_\mathcal{F}(Q^*) = 1 + O(\alpha_s^4)
\end{equation}
It remains an open question whether this result can be extended to 
higher orders or not.

Finally \cite{kat}, can this work be extended to include the
singlet axial-vector operator and hence the leading power in the
Ellis-Jaffe sum rule \cite{ej}? It seems not. The problem is 
\cite{kn} that the analogue of (\ref{c4}) involves a leading 
power with $J_{\mu 5}$ replaced in (\ref{c3}) by the gauge-dependent 
symmetry current of Adler \cite{adler1} and Bardeen \cite{bard}.
\vskip 4mm
\noindent
{\large\bf Acknowledgements}
\vspace{2mm}

I thank David Broadhurst, Stan Brodsky and Andrei Kataev for 
encouraging me to work on this subject again.

\end{document}